\documentclass[twocolumn,amsmath,amssymb,longbibliography,aps,prb]{revtex4-2}
\usepackage[dvipdfmx]{graphicx}

\usepackage{booktabs}
\usepackage{siunitx}
\usepackage{mathrsfs}
\usepackage{bm}
\usepackage{physics}
\usepackage{dcolumn}
\usepackage{epstopdf}
\usepackage{dsfont}
\usepackage{tabularx}
\usepackage{array}
\usepackage{float}
\usepackage{color}
\usepackage{epstopdf}
\usepackage{multirow}
\usepackage[T1]{fontenc}
\usepackage[colorlinks,linkcolor=blue,anchorcolor=blue,citecolor=blue,urlcolor=blue]{hyperref}
\providecommand{\U}[1]{\protect\rule{.1in}{.1in}}

\renewcommand{\Im}{\operatorname{Im}}

\usepackage{amsfonts}
\usepackage{autobreak}
\usepackage{xspace}
\usepackage{epstopdf}
\usepackage{dcolumn}
\usepackage{longtable}
\usepackage{subfigure}
\usepackage{multirow}
\usepackage{comment}
\usepackage{braket}
\usepackage{lipsum}
\usepackage{mathdots}
\usepackage[version=3]{mhchem} 
\usepackage{pifont}

\newcommand{\TMA}{\ce{Ti_2MnAl}}

\begin{document}

\title{Fermi arcs around magnetic domain walls \\ in a compensated ferrimagnetic Weyl semimetal \ce{Ti2MnAl} }

\newcommand{\authA}{
\author{Yuta Furusho} \thanks{furusho.yuta@mbp.phys.kyushu-u.ac.jp}}
\newcommand{\authB}{
\author{Tomonari Meguro} \thanks{meguro.tomonari@mbp.phys.kyushu-u.ac.jp}}
\newcommand{\authC}{
\author{Kentaro Nomura} \thanks{nomura.kentaro@phys.kyushu-u.ac.jp}}

\newcommand{\affiA}{Department of Physics, Kyushu University, Fukuoka 819-0395, Japan}

 \authA
 \authB
 \authC
 \affiliation{\affiA}

\newcommand{\abstbody}{
Fermi arcs are one of the characteristic features of Weyl semimetals, appearing as surface states that connect Weyl points with opposite chiralities. It has also been suggested that Fermi arcs can emerge in the bulk due to the interplay between magnetic textures and Weyl physics. We focus on \ce{Ti2MnAl} which is an ideal magnetic Weyl semimetal with a compensated ferrimagnetic order. We systematically analyze domain wall-induced Fermi arcs in \ce{Ti2MnAl} using an effective tight-binding model. By varying the strength of spin–orbit coupling, we confirmed that these domain wall-induced Fermi arcs emerge as a result of shifts in the positions of the Weyl points. Furthermore, we found that these domain wall-induced Fermi arcs in \ce{Ti2MnAl} originate from the Chern number and represent a topologically robust state that is independent of the domain wall width.
}

 \begin{abstract}
  \abstbody
 \end{abstract}

\maketitle
{\it Introduction.}---
Weyl semimetals (WSMs) are three-dimensional topological materials characterized by their band-touching points known as Weyl points~\cite{WSM_01,WSM_02,WSM_03,WSM_04}. These Weyl points act as sources or sinks of Berry curvature in momentum space and exhibit positive or negative chirality. A key feature of WSMs is the presence of distinctive surface states known as Fermi arcs that connect Weyl points with opposite chiralities and are topologically protected by nonzero Chern numbers~\cite{WSM_03,WSM_04}. In contrast to these conventional surface Fermi arcs, it has been reported that magnetic domain walls (DW) can also induce Fermi arc structures in the bulk~\cite{FA_DW_Araki,FA_DW_PRX1,FA_DW_PRX2}. Such DW Fermi arcs have been demonstrated in simple models, such as the Wilson-Dirac model, where they arise due to the shifts in the Weyl point positions accompanying changes in the magnetization direction~\cite{FA_DW_Araki_lattice}. However, in realistic materials these DW Fermi arcs have been calculated only in limited cases (e.g., in \ce{Mn3Sn})~\cite{FA_DW_Mn3Sn,FA_DW_Mn3Sn_2,FA_DW_Mn3Sn_3}, and systematic studies on their topological protection and correspondence with Weyl point shifts are still lacking.

Recent first-principles calculations have identified \ce{Ti2MnAl} as an ideal magnetic WSM~\cite{TMA_abini_0}. In this material, the Weyl points lie extremely close to the Fermi energy, differing by approximately $14$~meV. \ce{Ti2MnAl} is a compensated ferrimagnet with the inverse Heusler structure and both time-reversal $\mathcal{T}$ and parity  $\mathcal{P}$ symmetry are broken. Unlike regular Heusler structures, \ce{Ti2MnAl} exhibits Weyl points even in the absence of spin–orbit coupling (SOC). Although SOC in \ce{Ti2MnAl} is small, it plays a crucial role in reducing the symmetry of the system and shifting the positions of the Weyl points. 

In this study, we analyze the DW Fermi arcs using an effective model of \ce{Ti2MnAl}. First, we check that the Fermi arcs emerge in this model under open boundary conditions. Next, we investigate the DW states for the case of a single-layer DW. By varying the SOC strength, we verify that these states arise from the shift in the positions of the Weyl points. Furthermore, by calculating the Chern numbers, we demonstrate that the DW states are topologically protected—that is, they are DW Fermi arcs. Finally, by calculating the DW Fermi arcs as varying the DW width, we find that these states remain stable regardless of the DW width.
\begin{figure}[t]
\centering
\includegraphics[width=1.0\linewidth]{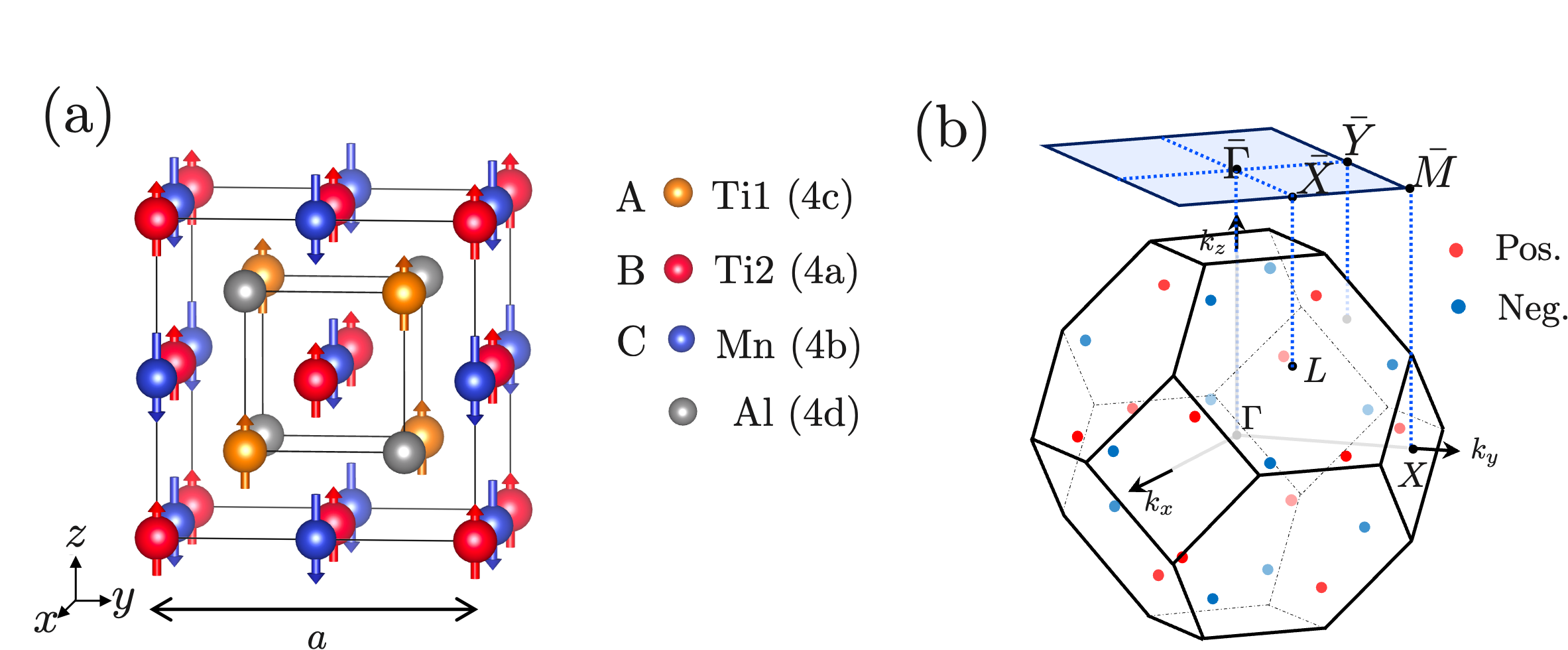}
\caption{(a) The structure of \TMA. $4a, 4b, 4c, 4d$ represents Wyckoff positions. Here, $a$ is the lattice constant of \ce{Ti2MnAl}. The crystal structure is illustrated by using the VESTA software~\cite{VESTA}. (b) The bulk Brillouin zone (BZ) and (001) surface BZ with high symmetry points of \ce{Ti2MnAl}. The colored points are 24 Weyl points with positive (red) and negative (blue) chirality.}
\label{structure}
\end{figure}

{\it Methods.}---
\ce{Ti_2MnAl} has an inverse Heusler structure (Space group No. 216; $F\bar{4}3m$) with four inequivalent Wyckoff positions (4a, 4b, 4c, and 4d), as shown in Fig.~\ref{structure}(a)~\cite{TMA_abini_0,TMA_abini_1,TMA_abini_2}. \ce{Ti} atoms are at 4a (0, 0, 0) and 4c (0.25, 0.25, 0.25), \ce{Mn} are at 4b (0.5, 0.5, 0.5) and \ce{Al} are at 4d (0.75, 0.75, 0.75) Wyckoff positions. 
Furthermore, \ce{Ti2MnAl} is a compensated ferrimagnet in which the overall magnetization cancels out due to the contributions of Ti and Mn.
In the absence of SOC, 24 Weyl points emerge in accordance with the space group symmetry. 
When SOC is present, the symmetry is reduced to reflect the magnetic structure, resulting in a shift in the positions of the Weyl points~\cite{TMA_abini_0}.

To investigate the Fermi arcs around a magnetic DW, we introduce the effective tight-binding model of \ce{Ti_2MnAl} proposed in the Ref. \ref{TMA66model}.
This model is a single-orbital $6\times6$ model constructed by excluding \ce{Al}, which is not related to the magnetism and the band structure around the Weyl points.
This model takes the crystal symmetry into account and reflects the 24 Weyl points of \ce{Ti_2MnAl} (Fig.~\ref{structure}(b)).
The Hamiltonian of this model is constructed from the hopping term $H_{\mathrm{t}}$, the exchange term $H_{\mathrm{exc}}$ and the SOC term $H_{\mathrm{soc}}$.
The explicit expression is given as follows:
\begin{align}\label{TMA_TB}
H 
=&H_{\mathrm{t}}+H_{\mathrm{exc}}+H_{\mathrm{soc}}\notag\\ 
=&\sum_{i\alpha}\sum_{s}\epsilon_{\alpha}c^{\dagger}_{i\alpha s}c_{i\alpha s}-\sum_{\braket{ij}\alpha\beta}\sum_{s}t_{\alpha\beta}c^{\dagger}_{i\alpha s}c_{j\beta s}\notag\\
&-\sum_{i\alpha}\sum_{ss^{\prime}}J_{\alpha}c^{\dagger}_{i\alpha s}(\hat{\bm{n}}\cdot\bm{\sigma})_{ss^{\prime}}c_{i\alpha s^{\prime}}\\
&+i\frac{2\lambda_{\mathrm{soc}}}{a^2}\sum_{\braket{ij}\alpha}\sum_{ss^{\prime}}c^{\dagger}_{i\alpha s}[(\bm{d}_{1}^{\alpha ij}\times\bm{d}_{2}^{\alpha ij})\cdot\bm{\sigma}]_{ss^{\prime}}c_{j\alpha s^{\prime}}\notag,
\end{align}
where $c^{\dagger}_{i\alpha s}$ denotes the electron creation operator at the $i$-th site with sublattice $\alpha = \text{A},\text{B},\text{C}$ (Ti1, Ti2, Mn) and spin $s=\uparrow,\downarrow$. 
The first term describes the on-site energy $\epsilon_{\alpha}$.
The second term is the nearest-neighbor hopping with an amplitude $t_{\alpha\beta}$. $\braket{ij}$ represents nearest neighbor hopping from site $i$ to $j$.
The third term represents the exchange coupling between conduction electron spin $\bm{\sigma}$ and mean field for compensated ferrimagnetic moments $\hat{\bm{n}}$ which is defined as $\hat{\bm{n}} = \bm{m}_{\mathrm{A}}/|\bm{m}_\mathrm{A}|=\bm{m}_{\mathrm{B}}/|\bm{m}_\mathrm{B}|=-\bm{m}_{\mathrm{C}}/|\bm{m}_\mathrm{C}|$ where $\bm{m}_{\alpha}$ denotes the magnetization of the $\alpha$ atom. $J_{\alpha}$ is the strength of exchange coupling.
The last term is the intrinsic SOC with a strength of $\lambda_{\mathrm{soc}}$. $\bm{d}_{1,2}^{\alpha ij}$ are the two nearest-neighbor vectors traversed between sites $i$ to $j$ of sublattice $\alpha$. For the calculations, periodic boundary conditions are imposed along the $x$ and $y$ directions, and the system is Fourier-transformed. For the $z$ direction, we use a real space coordinate to study the Fermi arcs. More details are provided in Appendix~\ref{app:A}.

To analyze the Fermi arcs, we employ the recursive Green's function method, which allows us for calculating the surface Green's function $G_{\mathrm{surf}}$ in a semi-infinite system. This is a widely used technique for calculating surface states~\cite{Recursive00,Recursive01,Recursive02}. Using the obtained Green's function $G_{\mathrm{surf}}$, we calculate the $\bm{k}$-resolved local density of states (LDOS) at the surface to extract the Fermi arcs:
\begin{align}
    {\mathrm{LDOS}}\ (E,\bm{k}) = -\frac{1}{\pi}\Im G_{\mathrm{surf}}(E+i0^{+},\bm{k}),
\label{LDOS_surf}
\end{align}
where $i0^{+}$ is a infinitesimal imaginary part and $\bm{k}=(k_x,k_y)$.
In this work, we apply this method to calculate the Fermi arcs at the DW~\cite{FA_DW_Mn3Sn}. 
For this purpose, we first consider the entire system with a DW, described by
\begin{align}
{H}_\mathrm{total} &= 
\begin{pmatrix}
    H_\mathrm{L}            & V_\mathrm{L}           & 0\\
    V_\mathrm{L}^{\dagger}  & H_\mathrm{C}           & V_\mathrm{R}\\
    0                       & V_\mathrm{R}^{\dagger} & H_\mathrm{R}
\end{pmatrix}.
\end{align}
Here, $H_{\mathrm{L}}$ and $H_{\mathrm{R}}$ represent the uniformly magnetized regions that sandwich the DW, while $H_{\mathrm{C}}$ describes the DW region. $V_{\mathrm{L/R}}$ represents the interaction between the left/right region and the DW region. The Green's function corresponding to $H_{\mathrm{C}}$ is given by
\begin{align}\label{DW_system}
G_{\mathrm{C}}(E,\bm{k}) = \frac{1}{E - H_{\mathrm{C}} - \Sigma_{\mathrm{R}} - \Sigma_{\mathrm{L}}},
\end{align}
where $\Sigma_{\mathrm{R/L}}$ denote the self-energies of the right and left regions, respectively. We use this Green's function $G_{\mathrm{C}}$ to calculate the LDOS at the DW. 
\begin{figure}[t]
\centering
\includegraphics[width=1.0\columnwidth]{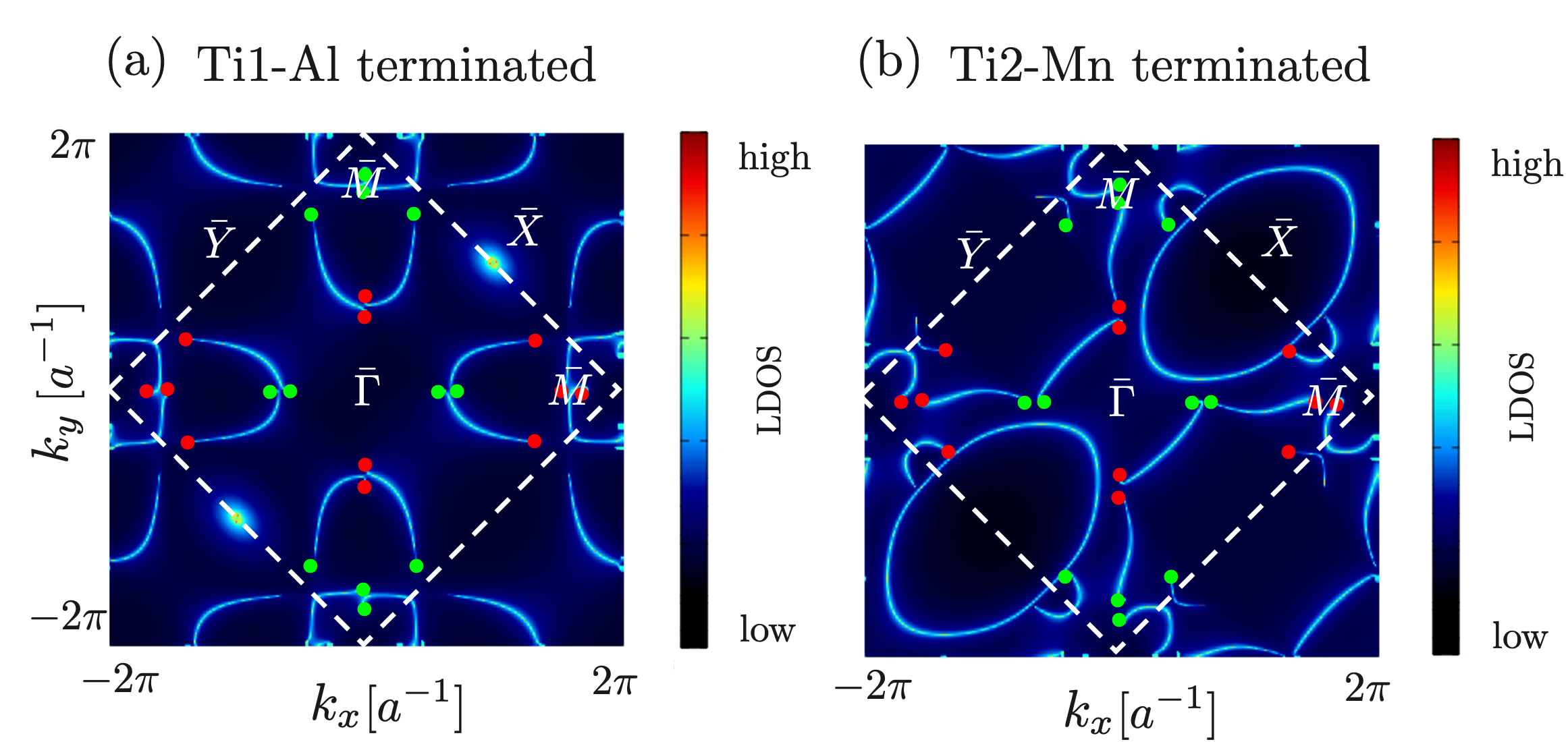}
\caption{
The (001) surface states for \ce{Ti2MnAl} at $E=E_{F}$ with magnetization oriented along the $z$-axis. Weyl points are denoted as green and red circles. (a) \ce{Ti}1-\ce{Al} terminated. (b) \ce{Ti}2-\ce{Mn} terminated.
}
\label{Fig:FA_OBC}
\end{figure}
\begin{figure}[t]
\centering
\includegraphics[width=1.0\columnwidth]{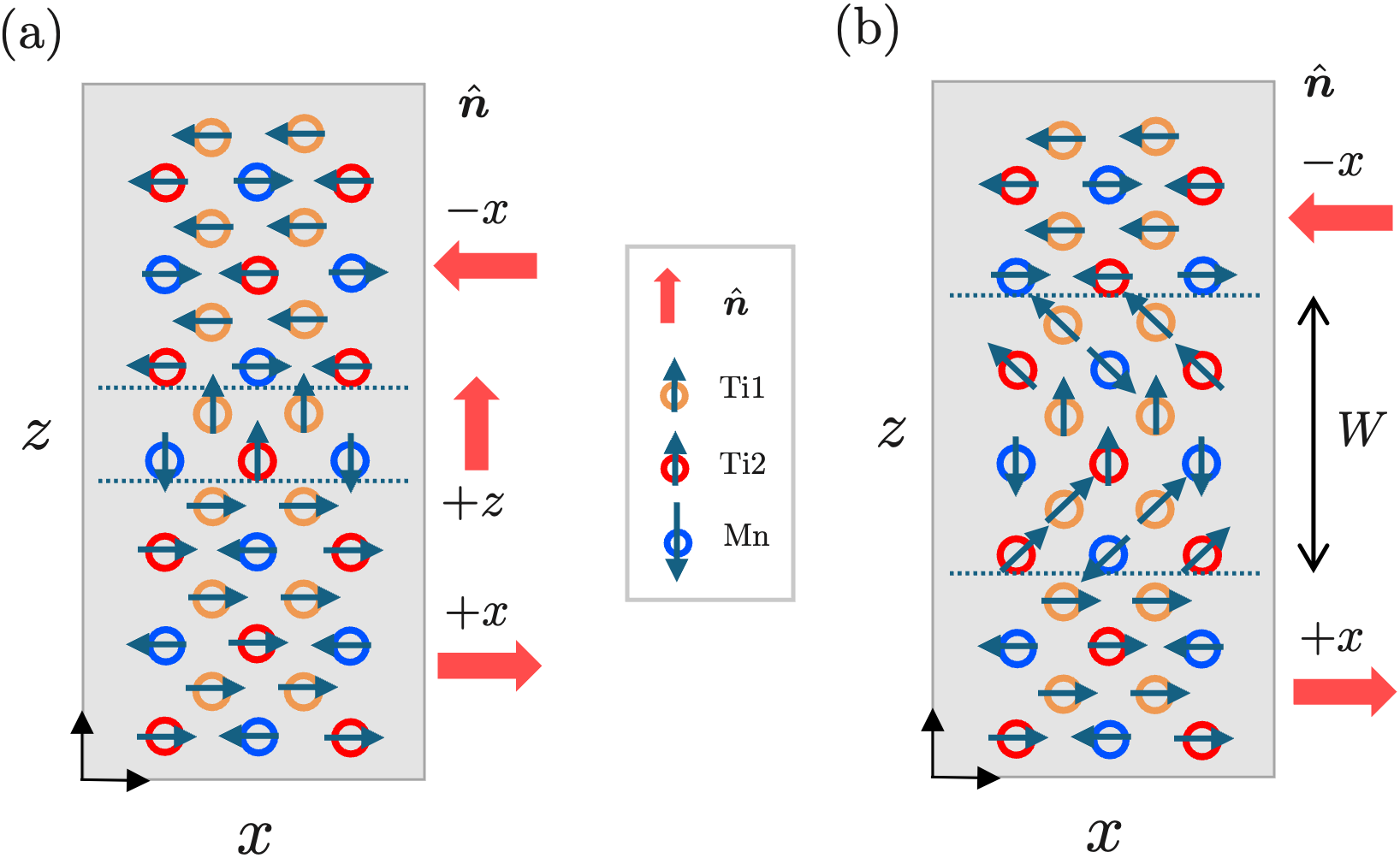}
\caption{\label{Schematic_DW}
Schematic illustrations of the system with a DW which is sandwiched between $+x$ and $-x$ uniform magnetization.
$\hat{\bm{n}}$ represents a compensated ferrimagnetic moment ($\hat{\bm{n}} = \bm{m}_{\mathrm{A}}/|\bm{m}_\mathrm{A}|=\bm{m}_{\mathrm{B}}/|\bm{m}_\mathrm{B}|=-\bm{m}_{\mathrm{C}}/|\bm{m}_\mathrm{C}|$).
(a) A system with a single-layer magnetic DW.
(b) A system with a DW of width $W$.
}
\label{Schematic_DW}
\end{figure}
\begin{figure}[H]
\centering
\includegraphics[width=1.0\columnwidth]{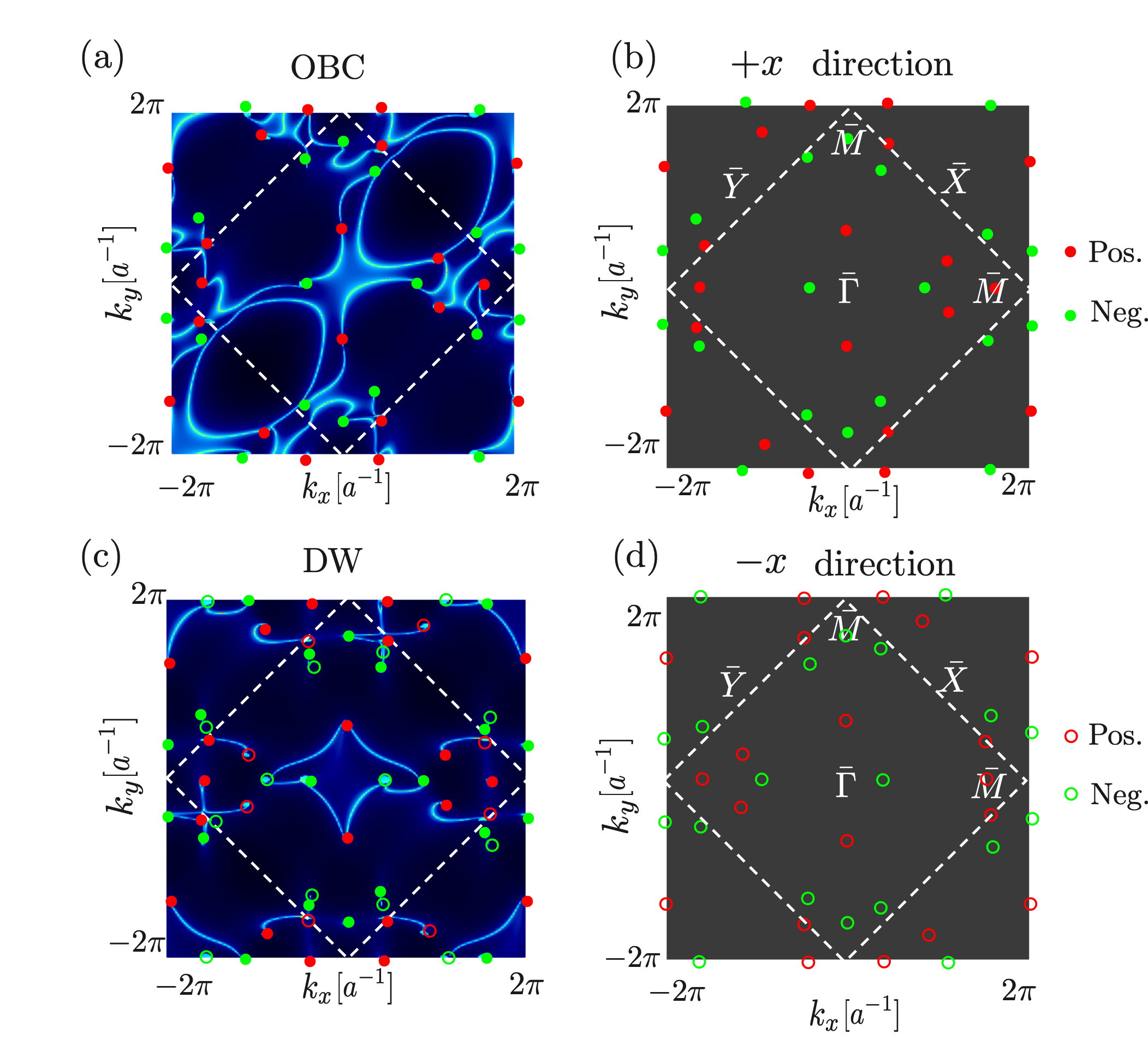}
\caption{ 
The left panels (a,c) are surface states and DW states at $E=E_F$. In each panel, the Weyl points are also plotted. The right panels (b,d) are schematic illustrations of the Weyl points with $+x$ and $-x$ magnetization, respectively. 
(a) The (001) surface states under OBC with $+x$ magnetization. (b) Weyl points with $+x$ magnetization. (c) The DW states sandwiched between regions with $+x$ and $-x$ magnetization. (d) Weyl points with $-x$ magnetization.
}
\label{Fig:Schematic_WPs}
\end{figure}
\begin{figure*}[t]
\centering
\includegraphics[width=1.5\columnwidth]{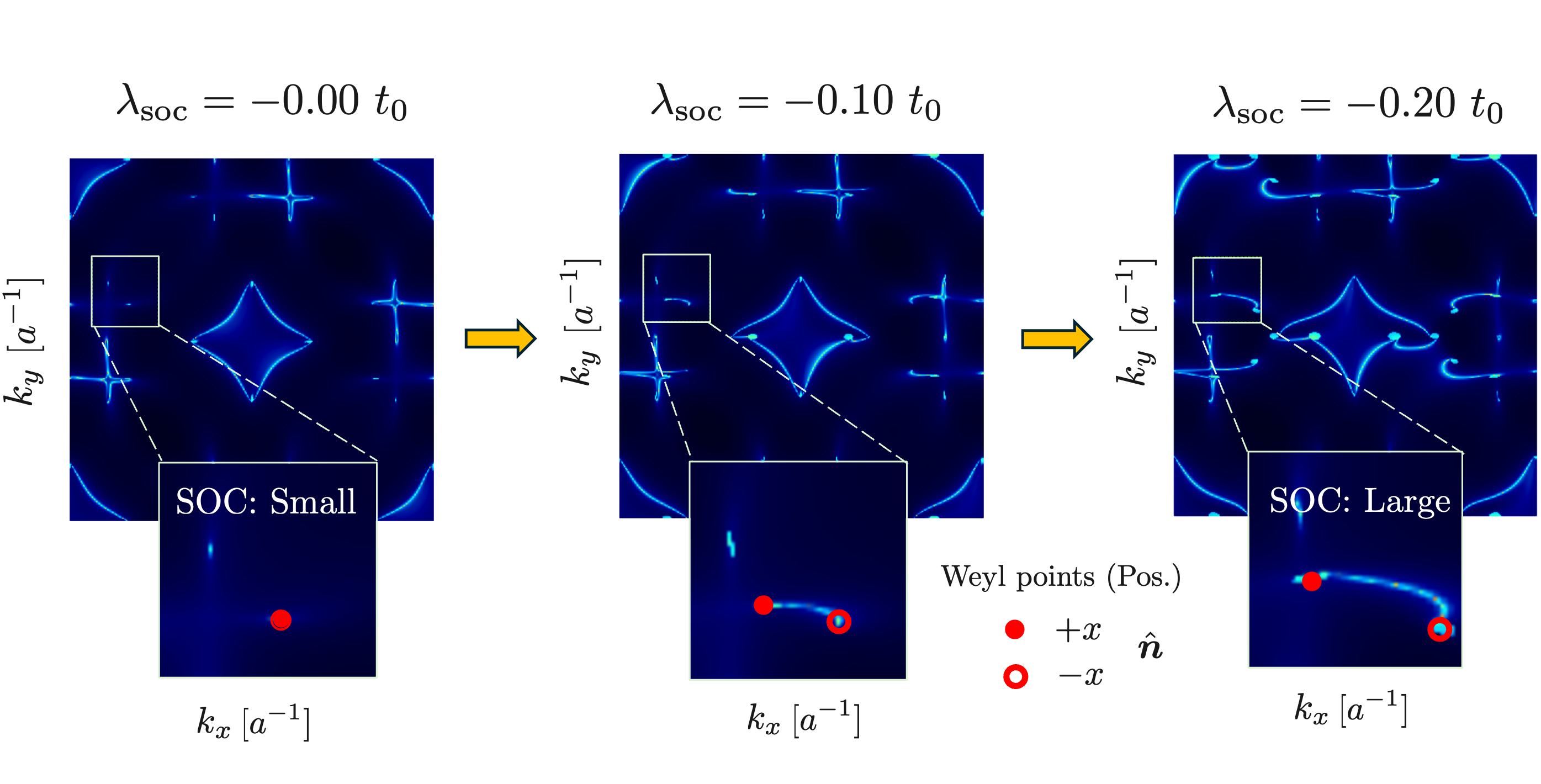}
\caption{
SOC strength dependence of the states at $E=E_F$ at the DW.
SOC becomes stronger toward the right. The red filled circle represents the Weyl point with positive chirality when the magnetization $\hat{\bm{n}}$ is oriented along the $+x$ direction, whereas the red hollow circle represents the Weyl point with positive chirality when the magnetization $\hat{\bm{n}}$ is oriented along the $-x$ direction.
}
\label{Fig:DW_SOC}
\end{figure*}
\begin{figure}[t]
\centering
\includegraphics[width=1.0\columnwidth]{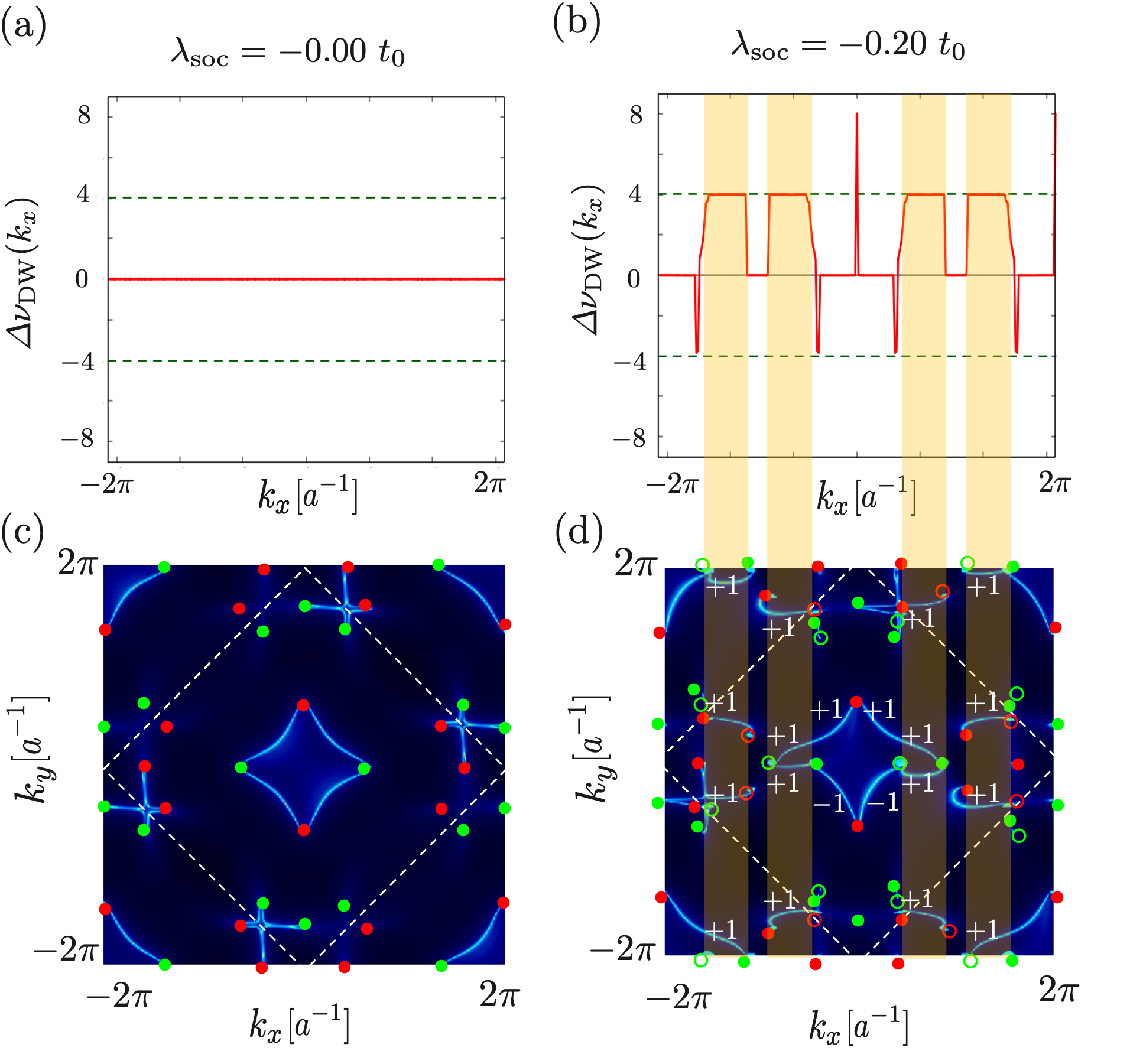}
\caption{
The top rows of panels (a,b) display $\mathrm{\varDelta\nu_{\mathrm{DW}}}$ for SOC strengths of $\lambda_{\mathrm{soc}}=-0.00\,t_{0}$ and $\lambda_{\mathrm{soc}}=-0.20\,t_{0}$, respectively. The bottom rows (c,d) show the energy states at $E=E_F$ at the DW corresponding to these panels.
}
\label{Fig:Chern_num}
\end{figure}
\begin{figure*}[t]
\centering
\includegraphics[width=1.6\columnwidth]{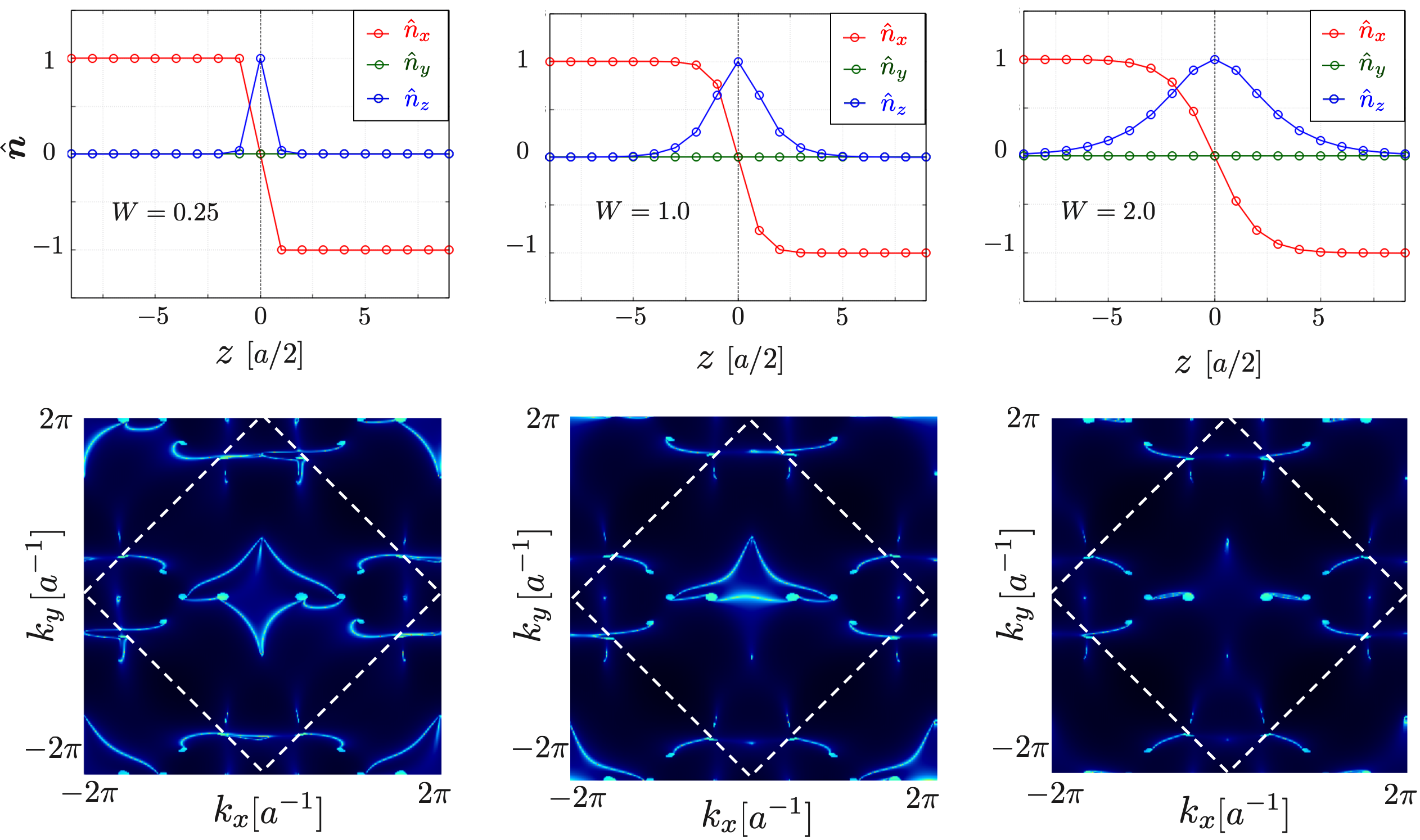}
\caption{
DW Fermi arcs when the DW width is varied. The top panels depict the orientation of the unit magnetization vector $\hat{\bm{n}}$, while the bottom panels display the corresponding states at the center of the DW for each case. $W$ [$a/2$] is the width of DW.
}
\label{Fig:DW_width_vary}
\end{figure*}
\begin{align}
    {\mathrm{LDOS}}\ (E,\bm{k}) = -\frac{1}{\pi}\Im G_{\mathrm{C}}(E+i0^{+},\bm{k}).
\label{LDOS_DW}
\end{align}
More details are provided in Appendix~\ref{app:B}. 

{\it Fermi arcs (OBC).}---
Under open boundary conditions (OBC), we confirm that Fermi arcs appear in this model. We perform calculations assuming that the magnetization is directed along the $z$-axis, as shown in Fig.~\ref{structure}(a). In this case, Eqs.~(\ref{G_top}),~(\ref{G_bottom}) and~(\ref{LDOS_surf}) are employed to calculate Green's function $G_{\mathrm{surf}}$ and LDOS at the surface. Figure~\ref{Fig:FA_OBC} shows the LDOS at the (001) surface at $E=E_{F}$. Here, $E_{F}$ denotes the bulk Fermi energy when the SOC is included. Two terminations are displayed in Fig.~\ref{Fig:FA_OBC}: (a) Ti1-Al terminated and (b) Ti2-Mn terminated. The green and red dots represent the Weyl points with negative and positive chirality, respectively. The Fermi arcs connecting the Weyl points with negative and positive chirality is observed, as expected under standard OBC~\cite{TMA_abini_0}. Furthermore, we also find that the Fermi arcs observed on Ti1-Al terminated and Ti2-Mn terminated surfaces differ substantially.

{\it Fermi arcs (DW).}---
To observe the DW states and confirm that they correspond to the DW Fermi arcs, we first consider the possible magnetic configurations. For this purpose, we take into account that the configuration of magnetic moments on Ti and Mn aligned in the opposite direction is stable~\cite{TMA66model}. Figure \ref{Schematic_DW} shows the two possible cases of magnetic DW textures. The DW is assumed to be sandwiched between regions with magnetization along the $+x$ and $-x$ directions. We first analyze the case where the magnetic DW consists of a single layer, as shown in Fig.~\ref{Schematic_DW}(a).

Before investigating the DW states, we confirm that the Weyl points and Fermi arcs appear as expected when the magnetization is aligned in the $+x$ direction. Figure~\ref{Fig:Schematic_WPs}(a) shows the LDOS on the (001) surface, and we used Eq.~(\ref{LDOS_surf}) to calculate the LDOS. Here, we simultaneously display the Weyl points for the $+x$-magnetized case (Fig.~\ref{Fig:Schematic_WPs}(b)). Figure~\ref{Fig:Schematic_WPs}(a) shows that the surface states connect the Weyl points with positive and negative chirality, confirming the presence of Fermi arcs. Next, we investigate the DW states in the presence of a DW. Figure~\ref{Fig:Schematic_WPs}(c) shows the LDOS at the DW, and we used Eq.~(\ref{LDOS_DW}) to calculate the LDOS. Here, we simultaneously plot the Weyl points for the $+x$-magnetized case (Fig.~\ref{Fig:Schematic_WPs}(b)) and the $-x$ magnetized case (Fig.~\ref{Fig:Schematic_WPs}(d)). Figure~\ref{Fig:Schematic_WPs}(c) clearly shows that the DW states appear as connections between the Weyl points of the $+x$ and $-x$ magnetized regions.

To verify these observed states at the DW originate from the positioning of the Weyl points, we vary the strength of SOC. Since SOC alters the positions of the Weyl points, it is expected that the DW states will evolve accordingly with increasing SOC strength. The results are shown in Fig.~\ref{Fig:DW_SOC}.
These results clearly show that the region of the LDOS connecting the Weyl points expands as the SOC strength increases. This behavior clearly demonstrates that it can be understood from the perspective of the shift of the Weyl points.

Next, to confirm these DW states represent topologically protected states referred to the DW Fermi arcs, we introduce the difference between the Chern numbers of two regions with $+x$ and $-x$ magnetization:
\begin{align}
    \mathrm{\varDelta\nu_{\mathrm{DW}}}(k_x)&=\nu_{+}(k_x)-\nu_{-}(k_x).
\label{Chern_num_DW}
\end{align}
$\nu_{\pm}(k_x)$ represents the Chern number for systems with uniform magnetization oriented along the $\pm x$ direction, and is defined as follows:
\begin{align}
    \nu_{\pm}(k_x)&=\frac{1 }{2\pi}\iint_{-2\pi/a}^{2\pi/a} b^{\pm}_x(k_x,k_y,k_z) dk_y dk_z ,
\label{Chern_num}
\end{align}
where $\bm{b}^{\pm}(\bm{k}) = \nabla_{\bm{k}}\times \bm{a}^{\pm}(\bm{k})$ is the Berry curvature and $\bm{a}^{\pm}(\bm{k}) = -i \sum_{E_{n\bm{k}}\leq E_{F}} \braket{n\bm{k}\pm|\nabla_{\bm{k}}|n\bm{k}\pm}$ is the Berry connection summed over states with energies below $E_F$. Here, $\ket{n\bm{k}\pm}$ represents the Bloch state for band $n$, wavevector $\bm{k}=(k_x,k_y)$ and magnetization $\pm x$. $E_{F}$ denotes the bulk Fermi energy when SOC is included. $a$ is the lattice constant.
Figure~\ref{Fig:Chern_num} shows the presence of states associated with finite $\mathrm{\varDelta\nu_{\mathrm{DW}}}$. Here, as shown in Fig.~\ref{Fig:Chern_num}(d), the numbers for each DW state are assigned so that their sum corresponds to $\mathrm{\varDelta\nu_{\mathrm{DW}}}$. This indicates that the DW states are topologically protected by $\mathrm{\varDelta\nu_{\mathrm{DW}}}$.

{\it Dependence of DW's width.}---
In practice, since DWs have a finite width, we investigate the effect of this width.
To examine the Fermi arcs as the width of the DW is varied, as shown in Fig.~\ref{Schematic_DW}(b), we vary the unit magnetization vector $\hat{\bm{n}}$ as follows:
\begin{align}
    \hat{\bm{n}}(z) &= \left(-\tanh(z/W),\ 0,\ \frac{1}{\cosh(z/W)}\right),
\label{mag_DW_width}
\end{align}
where $W$ denotes the width of the DW. Here, $z$ represents the position of the layers composed of Ti1, Ti2, and Mn, and the distance between adjacent layers is $a/2$. We examine the Fermi arcs by analyzing the LDOS at the center of the DW. Figure~\ref{Fig:DW_width_vary} shows that as the DW width increases, the $+1$ and $-1$ assigned in Fig.~\ref{Fig:Chern_num}(d) cancel each other out, leaving only regions with finite numbers. This indicates that the Fermi arcs induced by the DW are independent of the DW width and exist as topologically robust states.

{\it Conclusion.}---
In this study, we analyzed the DW Fermi arcs using an effective model of \ce{Ti2MnAl}. Our results indicate that the DW Fermi arcs can be understood in terms of the shifts in the positions of the Weyl points by varying the strength of SOC. These DW Fermi arcs are topologically protected states, as they are determined by the difference in the Chern numbers of the uniformly magnetized regions with opposite magnetization. Furthermore, they are independent of the DW width.

Although direct experimental observation of DW Fermi arcs is challenging, these states possess the unique ability to accumulate charge at the DW~\cite{FA_DW_Araki,FA_DW_Araki_lattice}. By using the fact that the Fermi arcs are independent of the DW width, variations in the charge density as changing the DW width can serve as an indirect signature of the DW Fermi arcs. A more detailed analysis of this possibility remains an important subject for future research. 

{\it Acknowledgments.}---
This work is supported by JST CREST, Grant Nos. JPMJCR18T2.

\appendix 

\section{Effective model in a slab geometry}\label{app:A}
In this Appendix A, we introduce the Fourier-transformed version of Eq. (\ref{TMA_TB}) in the $x,y$ direction. The magnetic DW is introduced by changing the magnetization vector $\hat{\bm{n}}$.
As shown in Eq.~(\ref{TMA_TB}), this model is composed of the hopping, the exchange interactions, and SOC. Upon performing the Fourier transformation, the following Hamiltonian is obtained. Here, we set the lattice constant $a = 1$ for simplicity. 

\noindent
Hopping term:
\begin{align} 
\mathcal{H}_\mathrm{t}(k_x,k_y) &= 
-\begin{pmatrix}
    H_\mathrm{t}^{0}        & V_\mathrm{t}           &                  &              &        & O\\
    V_\mathrm{t}^{\dagger}  & H_\mathrm{t}^{0}       & V_\mathrm{t}     &              &        &  \\
                            & V_\mathrm{t}^{\dagger} & H_\mathrm{t}^{0} & V_\mathrm{t} &        &  \\
                            &                        & \ddots           & \ddots       & \ddots &  \\
    O                       &                        &                  &              &        &
\end{pmatrix}.
\end{align}
Here, the diagonal part ${H}_\mathrm{t}^{0}$ and the off-diagonal part ${V}_\mathrm{t}$ are defined as,
\begin{align}
{H}_\mathrm{t}^{0} &= 
\begin{pmatrix}
    t_{\mathrm{BB}} f_{0} - \epsilon_\mathrm{B} & 2 t_{\mathrm{BC}} h_{0}                       & t_{\mathrm{AB}} g_{12}\\
    2 t_{\mathrm{BC}} h_{0}                     & t_{\mathrm{CC}} f_{0} - \epsilon_{\mathrm{C}} & t_{\mathrm{CA}} g_{11}\\
    t_{\mathrm{AB}} g_{12}                      & t_{\mathrm{CA}} g_{11}                        & t_{\mathrm{AA}} f_{0} - \epsilon_{\mathrm{A}}
\end{pmatrix},
\end{align}
\begin{align}
{V}_\mathrm{t} &= 
\begin{pmatrix}
    t_{\mathrm{BB}}f_{1}    & 2t_{\mathrm{BC}} h_{1}    & 0 \\
    2 t_{\mathrm{BC}} h_{1} & t_{\mathrm{CC}} f_{1}     & 0 \\
    t_{\mathrm{AB}}g_{11}   & t_{\mathrm{CA}}g_{12}     & t_{\mathrm{AA}}f_{1}
\end{pmatrix},
\end{align}
where
\begin{align*}
    f_{0} &= 4 \cos\frac{k_x}{2} \cos\frac{k_y}{2},\ \hspace{10pt}  
    f_{1} = 2 \left( \cos\frac{k_x}{2} + \cos\frac{k_y}{2} \right), \\
    g_{11} &= 2  \cos \left (\frac{k_x-k_y}{4}  \right),\ \hspace{2pt}    
    g_{12} = 2  \cos \left (\frac{k_x+k_y}{4}  \right), \\
    h_{0} &=  \cos\frac{k_x}{2} + \cos\frac{k_y}{2},\ \hspace{10pt}   
    h_{1} =  \frac{1}{2}.
\end{align*}
Here, $t_{ij}$ are hopping amplitude when hopping $i$-site to $j$-site and the various parameters are listed in Table~\ref{parameter_Hop}.

\begin{table*}[t]
    \centering
    \caption{Hopping parameters: $\epsilon_{\mathrm{A,B,C}}$ represents the on-site energies, and $t_{ij}$ ($i,j = \mathrm{A,B,C}$) denote the hopping amplitudes. The parameters are taken from the Ref.~\ref{TMA66model}. Here, $t_{0} = 0.42$ eV}
    \begin{tabular}{ccccccccc}\label{parameter_Hop}
    \\
    \toprule\toprule
         $\epsilon_{\mathrm{A}}$& 
         $\epsilon_{\mathrm{B}}$& 
         $\epsilon_{\mathrm{C}}$& 
         $t_{\mathrm{AB}}$& 
         $t_{\mathrm{BC}}$& 
         $t_{\mathrm{CA}}$& 
         $t_{\mathrm{AA}}$&
         $t_{\mathrm{BB}}$&
         $t_{\mathrm{CC}}$\\
    \midrule
         $-2.15\ t_{0}$& 
         $-2.15\ t_{0}$& 
         $-2.15\ t_{0}$& 
         $1.1\ t_{0}$&
         $0.4\ t_{0}$&
         $1.2\ t_{0}$&
         $0.05\ t_{0}$&
         $0.85\ t_{0}$&
         $-0.05\ t_{0}$\\
    \bottomrule\bottomrule
    \end{tabular}
\end{table*}

\begin{table*}[t]
    \centering
    \caption{Exchange \& SOC parameters: $J_{\mathrm{A,B,C}}$ represents the strength of exchange coupling, and $\lambda_{\mathrm{soc}}$ denotes the SOC hopping amplitude. The parameters are taken from the Ref.~\ref{TMA66model}. Here, $t_{0} = 0.42$ eV}
    \begin{tabular}{cccc}\label{parameter_Exchange_SOC}
    \\
    \toprule\toprule
         $J_{\mathrm{A}}$&
         $J_{\mathrm{B}}$&
         $J_{\mathrm{C}}$&
         $\lambda_{\mathrm{soc}}$\\
    \midrule
         $0.7\ t_{0}$&
         $0.7\ t_{0}$&
         $-1.7\ t_{0}$&
         $-0.2\ t_{0}$\\
    \bottomrule\bottomrule
    \end{tabular}
\end{table*}

\noindent
Exchange term:
\begin{align}
\mathcal{H}_\mathrm{exc}(k_x,k_y) &= 
\begin{pmatrix}
    H_\mathrm{exc}^{0}   &                    &        & O\\
                         & H_\mathrm{exc}^{0} &        &  \\
                         &                    & \ddots &  \\
    O                    &                    &        &
\end{pmatrix}.
\end{align}
Here, the diagonal part ${H}_\mathrm{exc}^{0}$ is defined as,
\begin{align}
{H}_\mathrm{exc}^{0} &= -
\begin{pmatrix}
    J_\mathrm{B}\hat{\bm{n}}\cdot\bm{\sigma} & 0                                  & 0 \\
    0                                   & J_\mathrm{C}\hat{\bm{n}}\cdot\bm{\sigma} & 0 \\
    0                                   & 0                                  & J_\mathrm{A}\hat{\bm{n}}\cdot\bm{\sigma}
\end{pmatrix}.
\end{align}
The various parameters are listed in Table~\ref{parameter_Exchange_SOC}.

\noindent
SOC term:
\begin{align}
\mathcal{H}_\mathrm{soc}(k_x,k_y) &= 
\begin{pmatrix}
    H_\mathrm{soc}^{0}       & V_\mathrm{soc}           &                    &                &        & O\\
    V_\mathrm{soc}^{\dagger} & H_\mathrm{soc}^{0}       & V_\mathrm{soc}     &                &        &  \\
                             & V_\mathrm{soc}^{\dagger} & H_\mathrm{soc}^{0} & V_\mathrm{soc} &        &  \\
                             &                          & \ddots             & \ddots         & \ddots &  \\
    O                        &                          &                    &                &        &  \\
\end{pmatrix}.
\end{align}
Here, the diagonal part ${H}_\mathrm{soc}^{0}$ and the off-diagonal part ${V}_\mathrm{soc}$ are defined as,
\begin{align}
{H}_\mathrm{soc}^{0} &= 
\begin{pmatrix}
    \lambda_\mathrm{soc}\bm{R}_{0}\cdot\bm{\sigma}  & 0                                               & 0\\  
    0                                               & -\lambda_\mathrm{soc}\bm{R}_{0}\cdot\bm{\sigma} & 0\\
    0                                               & 0                                               & 0
\end{pmatrix},
\end{align}
\begin{align}
{V}_\mathrm{soc} &=
\begin{pmatrix}
    \lambda_\mathrm{soc}\bm{R}_{1}\cdot\bm{\sigma} & 0                                               & 0\\
    0                                              & -\lambda_\mathrm{soc}\bm{R}_{1}\cdot\bm{\sigma} & 0\\
    0                                              & 0                                               & 0
\end{pmatrix},
\end{align}
where
\begin{align*}
    \bm{R}_{0} &= \left(\sin{\frac{k_x}{2}}\cos{\frac{k_y}{2}},\ -\cos{\frac{k_x}{2}}\sin{\frac{k_y}{2}},\ 0\right), \\
    \bm{R}_{1} &=\frac{1}{2} \left(-\sin{\frac{k_x}{2}},\ \sin{\frac{k_y}{2}},\ -i\cos{\frac{k_x}{2}}+i\cos{\frac{k_y}{2}}\right).
\end{align*}
The various parameters are listed in Table~\ref{parameter_Exchange_SOC}.

\section{Iterative Green's function method}\label{app:B}

\begin{figure*}[t]
    \centering
    \includegraphics[width=2.0\columnwidth]{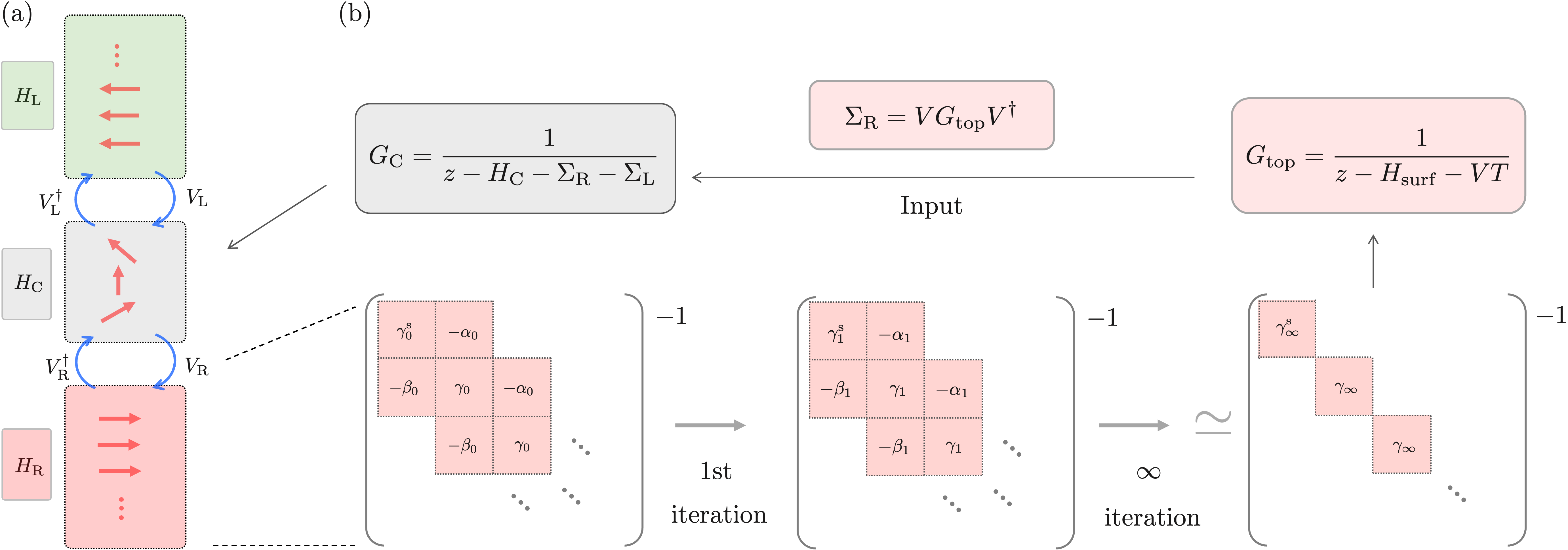}
    \caption{(a) Total system which a DW is introduced. $H_\mathrm{L}$ and $H_{\mathrm{R}}$ are semi-infinite systems with a uniform magnetization $\hat{\bm{n}}$. The DW is represented by the magnetization $\hat{\bm{n}}$ in $H_{\mathrm{C}}$ (b) Flow of the algorithm for calculating the center Green's function $G_{\mathrm{C}}$. Using iterative Green's function, we get the top Green's function $G_{\mathrm{top}}$ and construct the self-energy $\Sigma_{\mathrm{R}}$. Then we can calculate $G_{\mathrm{C}}$ by using $\Sigma_{\mathrm{R}}$ (and $\Sigma_{\mathrm{L}}$).
    }\label{Flow_GreenFunc}
\end{figure*}
In this section we introduce a iterative Green's function method and explain how to obtain the Green's function for a DW. This method allows for the calculation of the Green's function in semi-infinite systems and is generally used to compute the surface Green's functions, which can then be employed to visualize Fermi arcs. Here, we apply this method to compute the Green's function for a DW~\cite{FA_DW_Mn3Sn}. To analyze the DW, we consider the system shown in Fig.~\ref{Flow_GreenFunc}(a). Below, we assume a general form and consider the total Hamiltonian $H_{\mathrm{total}}$:
\begin{align}
{H}_\mathrm{total} &= 
\begin{pmatrix}
    H_\mathrm{L}            & V_\mathrm{L}           & 0\\
    V_\mathrm{L}^{\dagger}  & H_\mathrm{C}           & V_\mathrm{R}\\
    0                       & V_\mathrm{R}^{\dagger} & H_\mathrm{R}
\end{pmatrix},
\end{align}
where $H_{\mathrm{R}}$ and $H_{\mathrm{L}}$ represent the semi-infinite systems with uniform magnetization, while $H_{\mathrm{C}}$
constitutes the DW. Therefore, to calculate the Green's function for the DW, it is necessary to compute the Green's function corresponding to $H_{\mathrm{C}}$. The total Green's function is given by the following expression:
\begin{align}
{G}_\mathrm{total} &= 
\begin{pmatrix}
    G_\mathrm{L}  & G_\mathrm{LC} & G_\mathrm{LR}\\
    G_\mathrm{CL} & G_\mathrm{C}  & G_\mathrm{CR}\\
    G_\mathrm{RL} & G_\mathrm{RC} & G_\mathrm{R}
\end{pmatrix}.
\end{align}
The explicit expression for $G_{\mathrm{total}}$ is calculated using the following equation:
\begin{align}
(z-{H}_\mathrm{total}) G_\mathrm{total} = I,
\end{align}
where $z = (E + i0^{+})I$, $E$ is the energy, $0^{+}$ is a small positive quantity and $I$ is the identity matrix.
\begin{align} \label{CenterGreenFunc}
    G_\mathrm{C} = [z-H_\mathrm{C}-\Sigma_\mathrm{R} - \Sigma_\mathrm{L}]^{-1},
\end{align}
where
\begin{align} \label{SelfEnergy}
    \Sigma_\mathrm{R} &= V_\mathrm{R} (z-H_\mathrm{R})^{-1} V_\mathrm{R}^{\dagger}, \notag\\
    \Sigma_\mathrm{L} &= V_\mathrm{L}^{\dagger} (z-H_\mathrm{L})^{-1} V_\mathrm{L}.
\end{align}
Since $H_{\mathrm{R}}$ and $H_{\mathrm{L}}$ represent semi-infinite systems, it is not possible to directly compute $(z-H_R)^{-1}$ and $(z-H_R)^{-1}$ (note: $(z-H_{\mathrm{R/L}})^{-1}\neq G_{\mathrm{R/L}}$). To address this issue, we employ the iterative Green's function method~\cite{Recursive00,Recursive01,Recursive02}. Here, $H_{\mathrm{R}}$ and $H_{\mathrm{L}}$ are specifically assumed to have the following forms:
\begin{align}
{H}_\mathrm{R} &= 
\begin{pmatrix}
    {H}_\mathrm{surf}  & V             &        &        &        & O\\
    V^{\dagger}     & H             & V      &        &        &  \\
                    & V^{\dagger}   & H      & V      &        &  \\
                    &               & \ddots & \ddots & \ddots &  \\
    O               &               &        &        &        &
\end{pmatrix},
\end{align}
\begin{align}
{H}_\mathrm{L} &= 
\begin{pmatrix}
                &           &               &               &                & O\\
                &\ddots     & \ddots        & \ddots        &                &  \\
                &           & V^{\dagger}   & H             & V              &  \\
                &           &               & V^{\dagger}   & H              & V\\
    O           &           &               & & V^{\dagger} & H_{\mathrm{surf}} &
\end{pmatrix}.
\end{align}
In the \ce{Ti_2MnAl} model, $H\ (= H_{\mathrm{surf}})$ corresponds to $H_{\mathrm{t}}+H_{\mathrm{exc}}+ H_{\mathrm{soc}}$, and $V$ corresponds to $V_{\mathrm{t}}+V_{\mathrm{soc}}$ (see Appendix~\ref{app:A}). They satisfy the following equations:
\begin{align}
(z-{H}_\mathrm{R}) G = I, \label{Req}\\
(z-{H}_\mathrm{L}) \bar{G} = I, \label{Leq}
\end{align}
where
\begin{align}
G &= 
\begin{pmatrix}
    G_{00}      & G_{01}      & \cdots & G_{0\infty}    \\
    G_{10}      & G_{11}      &        & G_{1\infty}    \\
    \vdots      &             & \ddots & \vdots         \\
    G_{\infty0} & G_{\infty1} & \cdots & G_{\infty\infty}
\end{pmatrix},
\end{align}

\begin{align}
\bar{G} &= 
\begin{pmatrix}
    \bar{G}_{\infty\infty} & \cdots & \bar{G}_{\infty1} & \bar{G}_{\infty0} \\
    \vdots                 & \ddots &                   & \vdots            \\
    \bar{G}_{1\infty}      &        & \bar{G}_{11}      & \bar{G}_{10}      \\
    \bar{G}_{0\infty}      & \cdots & \bar{G}_{01}      & \bar{G}_{00}
\end{pmatrix}.
\end{align}
First, considering the Green's function for $H_{\mathrm{R}}$, using Eq. (\ref{Req}) we obtain the following equations:
\begin{align}
    &\gamma_0^{\mathrm{s}} G_{00} = I + \alpha_{0}G_{10}, \\
    &\gamma_0 G_{n,0}             = \alpha_{0}G_{n+1,0} + \beta_{0}G_{n-1,0} \ (n\geq 1).
\end{align}
where
\begin{align} 
    \alpha_{0} &= V, \ \ \beta_{0} = V^{\dagger}, \label{apb_a0}\\
    \gamma_{0}^{\mathrm{s}} &= z - H_{\mathrm{surf}}, \ \ \ \gamma_{0} = z - H. \label{apb_g0}
\end{align}
Here, the odd-numbered $G_{2n-1}$ ($n=1,2,\cdots$) can be eliminated, resulting in the following equations:
\begin{align*}
    &\gamma_1^{\mathrm{s}} G_{00} = I + \alpha_{1}G_{2,0}, \\
    &\gamma_1 G_{2n,0}        = \alpha_{1}G_{2(n+1),0} + \beta_{1}G_{2(n-1),0} \ (n\geq 1),
\end{align*}
where
\begin{align*}\label{apb_a1}
    \alpha_{1}              =& \alpha_{0}(\gamma_{0})^{-1}\alpha_{0}, \\
    \stepcounter{equation}\tag{\theequation}
    \beta_{1}               =& \beta_{0}(\gamma_{0})^{-1}\beta_{0}, \\
    \gamma_{1}^{\mathrm{s}} =& \gamma_{0}^{\mathrm{s}} - \alpha_{0}(\gamma_{0})^{-1}\beta_{0}, \\
    \gamma_{1}              =& \gamma_{0} - \alpha_{0}(\gamma_{0})^{-1}\beta_{0} \\
                             &\ \ \ \ \ \ \ - \beta_{0}(\gamma_{0})^{-1}\alpha_{0}.
\end{align*}
Repeating this operation $k$-times yields the following expressions:
\begin{align*}
    &\gamma_k^{\mathrm{s}} G_{00} = I + \alpha_{k}G_{2^{k},0}, \\
    &\gamma_k G_{2^{k}n,0}        = \alpha_{k}G_{2^{k}(n+1),0} + \beta_{k}G_{2^{k}(n-1),0} \ (n\geq 1),
\end{align*}
where
\begin{align}
    \alpha_{k}              =& \alpha_{k-1}(\gamma_{k-1})^{-1}\alpha_{k-1}, \\
    \stepcounter{equation}\tag{\theequation}
    \beta_{k}               =& \beta_{k-1}(\gamma_{k-1})^{-1}\beta_{k-1}, \\
    \gamma_{k}^{\mathrm{s}} =& \gamma_{k-1}^{\mathrm{s}} - \alpha_{k-1}(\gamma_{k-1})^{-1}\beta_{k-1}, \\
    \gamma_{k}              =& \gamma_{k-1} - \alpha_{k-1}(\gamma_{k-1})^{-1}\beta_{k-1} \\
                             &\ \ \ \ \ \ \ - \beta_{k-1}(\gamma_{k-1})^{-1}\alpha_{k-1}.
\label{apb_ak}
\end{align}
Next, we introduce
\begin{align}
    t_{k} = (\gamma_{k})^{-1} \beta_{k}, \ \ \ \ 
    \tilde{t}_{k} = (\gamma_{k})^{-1} \alpha_{k}.
\label{apb_t_k}
\end{align}
 For $k > 1$, we rewrite Eq.~(\ref{apb_ak}) as follows:
\begin{align*}
    \alpha_{k} =& \alpha_{0}\prod_{m=0}^{k-1}\tilde{t}_{m} ,\ \ \ \ 
    \beta_{k}  = \beta_{0} \prod_{m=0}^{k-1}{t}_{m}, \\
    \gamma_{k}^{\mathrm{s}} 
                            =& z - H_{\mathrm{surf}} - VT_{k}, \\
    \gamma_{k} 
               =& \gamma_{0} - \sum\limits_{n=1}^{k} \alpha_{n-1} t_{n-1} - \sum\limits_{n=1}^{k} \beta_{n-1} \tilde{t}_{n-1} \\
               =& 
                 \gamma_{0} - \alpha_{0} t_{0} - \beta_{0} t_{0} \\
                  &- \alpha_{0} \sum_{n=1}^{k-1} \left(\prod_{m=0}^{n-1}\tilde{t}_{m}\right) t_{n} - \beta_{0} \sum_{n=1}^{k-1} \left(\prod_{m=0}^{n-1}{t}_{m}\right) \tilde{t}_{n} \\
               \stepcounter{equation}\tag{\theequation}
               =& z - H - VT_{k} - V^{\dagger} \tilde{T}_{k}, \\
\end{align*}
where we use Eq. (\ref{apb_a0}) and Eq. (\ref{apb_g0}). $T_k$ and $\tilde{T}_{k}$ are defined as follows:
\begin{align*}
    T_{k}         &= t_{0}         + \sum_{n=1}^{k-1} \left(\prod_{m=0}^{n-1}\tilde{t}_{m}\right) t_{n} 
                   = t_{0} + \tilde{t}_{0}t_{1} + \cdots, \\
    \tilde{T}_{k} &= \tilde{t}_{0} + \sum_{n=1}^{k-1} \left(\prod_{m=0}^{n-1}{t}_{m}\right) \tilde{t}_{n}
                   = \tilde{t}_{0} + t_{0}\tilde{t}_{1} + \cdots,
\end{align*}
where we use Eq.~(\ref{apb_ak}). Substitute Eq.~(\ref{apb_ak}) into Eq.~(\ref{apb_t_k}), then
\begin{align*}
    t_{k}         &=  \left(I - t_{k-1}\tilde{t}_{k-1} - \tilde{t}_{k-1}t_{k-1}\right)^{-1} t_{k-1}^{2}, \\
    \tilde{t}_{k} &=  \left(I - t_{k-1}\tilde{t}_{k-1} - \tilde{t}_{k-1}t_{k-1}\right)^{-1} \tilde{t}_{k-1}^{2}.
\end{align*}
We can calculate $t_{k}$ and $\tilde{t}_{k}$ instead of $\alpha_{k}$, $\beta_{k}$.
Taking $k \to \infty$, the T-matrix is obtained.
\begin{align}
    T = \lim_{k \to \infty} T_{k}, \ \ \ \ 
    \tilde{T} = \lim_{k \to \infty} \tilde{T}_{k}.
\end{align}
When $k \to \infty$, as shown in Fig.~(\ref{Flow_GreenFunc}), the off-diagonal elements become negligible, and the surface Green's function for $H_{\mathrm{R}}$ is obtained by ${\gamma_{\infty}^{s}}^{-1}$. By applying the same procedure to $H_{\mathrm{L}}$, the surface Green's function is obtained as follows:
\begin{align}
    {G}_{00} &= (z-H_{\mathrm{surf}}-VT)^{-1} = G_{\mathrm{top}}, \label{G_top}\\
    \bar{G}_{00} &= (z-H_{\mathrm{surf}}-V^{\dagger}\tilde{T})^{-1} = G_{\mathrm{bottom}}. \label{G_bottom}
\end{align}
Then Eq.~(\ref{SelfEnergy}) becomes
\begin{align}
    \Sigma_\mathrm{R} &= V_\mathrm{R} (z-H_\mathrm{R})^{-1} V_\mathrm{R}^{\dagger}= V {G}_{00}  V^{\dagger}, \\
    \Sigma_\mathrm{L} &= V_\mathrm{L}^{\dagger} (z-H_\mathrm{L})^{-1} V_\mathrm{L} = V^{\dagger} \bar{G}_{00} V.
\end{align}
As we can see Eq. (\ref{CenterGreenFunc}), the Green's function at the magnetic DW can be calculated (see also Fig. \ref{Flow_GreenFunc}).


\end{document}